\documentclass[%
 reprint,
 nofootinbib,
 amsmath,amssymb,
 aps,
]{revtex4-2}

\usepackage{braket}
\usepackage{graphicx}
\usepackage{dcolumn}
\usepackage{bm}
\usepackage{amsmath}
\usepackage{xcolor}
\usepackage{leftindex}

\begin{document}

\title{Partial Quantisation of Non-Hermitian Berry Phases in Time-Varying Media}

\author{Calvin M. Hooper}
\email{ch1122@exeter.ac.uk}
\affiliation{
 School of Physics and Astronomy, University of Exeter, Stocker Road, Exeter, EX4 4QL, UK
}


\begin{abstract}
A fundamental symmetry of the non-Hermitian operators describing wave-propagation in time-varying media imbue such systems with non-trivial topology. This topology may be measured directly in a wide range of experimental settings as a quantised real part of the Berry phase, contrasting unconstrained geometric gain or loss. This topological index is provided explicitly for practical examples, including a non-Hermitian analogue of the Su-Schrieffer-Heeger model.
\end{abstract}

\maketitle

%
%
As a wave propagates through a material, sudden changes to material properties (faster than a period of the wave) affect it in ways impossible in slowly-varying media~\cite{Galiffi2022}, realising temporal reflection~\cite{Bacot2016, Vezzoli2018}, frequency conversion~\cite{Horsley2023BlackHole, PachecoPea2024} or amplification~\cite{Wang2024, GanforninaAndrades2024}. Mathematically, this complexity is succinctly encoded by promoting scalar quantities such as the wavenumber to operators encoding coupling between different frequencies~\cite{Horsley2023Operators}.

While the addition of operators results in all but the simplest problems becoming analytically intractable~\cite{Kovacic1986, Lax2006}, this field is hardly the first to experience the difficulties of operator differential equations. Indeed, operators are the bane of many an undergraduate's introduction to quantum mechanics~\cite{Singh2001}, where gaps in intuition host phenomena from the quantum Hall effect~\cite{vonKlitzing2005, Tong2016} to many-body scarring~\cite{Moudgalya2022, Chandran2023}.

It thus seems natural to draw analogy between the study of classical waves in time-varying media, and quantum mechanics. Even in static media, such analogies form a basis for optical perturbation theory~\cite{Testorf1999, Remo1978}. However, time-varying media strengthen this link hugely, naturally realising~\cite{Hooper2025} effects equivalent to \(\mathcal{PT}\)-symmetry~\cite{Bender1998, Bender2007}, or being tuned to study field theories with synthetic dimensions~\cite{Cheng2025, Yuan2018}.

However, overemphasising this link hides an important difference between the operators of time-varying media and those of quantum mechanics: the former are typically non-Hermitian. Such non-Hermitian operators naturally exhibit phenomena neglected in standard quantum mechanics, but discussed frequently in its non-Hermitian generalisations~\cite{Ashida2020, Bender1998}. Most notable of these effects, and the focus of this work, is the emergence of defective (non-diagonalisable) operators~\cite{Nering1970}.

In this letter, we study topology in the inherently non-Hermitian setting of time-varying media, where boundaries between adiabatically disconnected regions are no longer demarcated solely by degenerate operators, but also by defective operators~\cite{Bergholtz2021} (serving intuitively as a non-Hermitian generalisation of the former). However, adiabatically impassable boundaries provide topological results only if they cannot be circumvented. In Hermitian settings, such circumvention is prevented by imposing additional symmetries~\cite{Ryu2010}, with experimental observation of topological effects being limited by how accurately these symmetries are realised.

\begin{figure}
    \centering
    \includegraphics[width=0.8\linewidth]{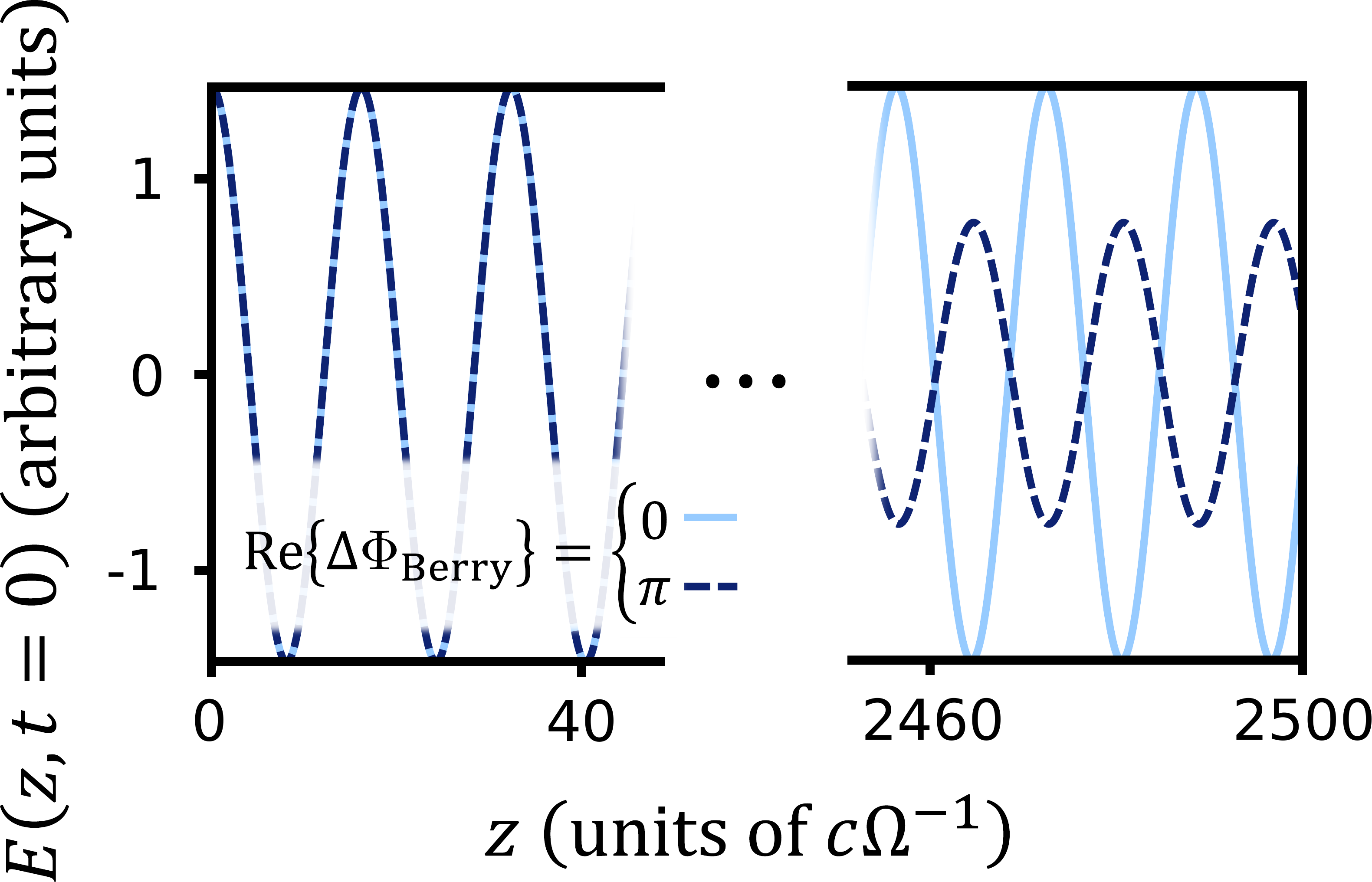}
    \caption{
        \textbf{Waves exhibiting different ``partially quantised'' complex Berry phases:}
        Between two ends of a long time-varying cavity, waves may pick up a geometric phase (denoted \({\rm Re}\left\{\Delta\Phi_{\rm Berry}\right\}\)) of either \(\pi\) (dashed) or \(0\) (undashed). In addition, such waves may experience an unquantised degree of geometric loss or gain, corresponding to a change in amplitude between ends of the cavity. The particular model used is the same as in Fig. \ref{fig:ConvergenceDemo}.
    }
    \label{fig:BerryPhaseDemo}
\end{figure}

Symmetry is a similarly key component of topology in non-Hermitian settings~\cite{Kawabata2026}, and our study is motivated by an inherent symmetry of such media: that they interact with real-valued waves equips them with a robust symmetry~\cite{Hooper2025, Hooper2026}. Whilst this symmetry is spontaneously broken in static media, time-varying media allow it to remain unbroken. In this letter, we demonstrate that this unbroken symmetry equips time-varying media with a robust and non-trivial non-Hermitian topology, observed via a quantisation in the real part of a complex Berry phase~\cite{Berry1984} (Fig. \ref{fig:BerryPhaseDemo}).

Furthermore, in constructing toy models to demonstrate this phenomenon,  we will also provide, as a corollary, an alternative proof of Zak phase quantisation in the SSH model~\cite{Asboth2016, Zak1989, Su1979}, and the \(\pi\) phase shifts when encircling Dirac points in 2D materials~\cite{CastroNeto2009, Ghahari2017, Breach2024}.

As a sufficient example to demonstrate the inherent topology of time-varying media, and to prove the partial quantisation condition in general, we consider the Berry phase~\cite{Berry1984, Asboth2016} evolved between the ends of a long time-varying slab with some slow spatial variation. In contrast to static media, such systems no longer preserve the frequencies of input waves. Thus, the Fourier Transform no longer converts the \(1+1\)-dimensional wave equation to a scalar differential equation. However, if one carries out the Fourier Transform regardless, one obtains~\cite{Horsley2023Operators} the operator-valued differential equation
\begin{equation}
    \frac{{\rm d}^2\widetilde{E}\left(z\right)}{{\rm d} z^2}=-\widehat{K}^2\left(z/L\right)\widetilde{E}\left(z\right),
    \label{eq:OperatorWaveEquation}
\end{equation}
where \(\widetilde{E}\left(z\right)\) is to be understood as an infinite-dimensional vector containing the frequency spectrum of the electric field at a 1-dimensional position \(z\). \(\widehat{K}\left(z/L\right)\) is then an operator generalisation of the wavenumber acting on this vector to convert between different frequencies as waves propagate through space. It is taken to be a (differentiable) function of \(z/L\), thus describing a profile of material parameters stretched over \(z\in \left[0,L\right]\) (focusing particularly on the adiabatic limit of \(L\to\infty\)). Finally, \(\widehat{K}\), is taken as a constant on either side of this range.

We now derive the key result of this paper: that topology arises from the fact that \(\widetilde{E}\left(z\right)\) is the Fourier Transform of a real-valued wave, implying \(\widetilde{E}\left(\omega\right)=\widetilde{E}^*\left(-\omega\right)\). In the language of \(\mathcal{PT}\)-symmetry~\cite{Bender1998, Bender2007}, this may be understood as a composite symmetry comprised of two separate operations: frequency reflection \(\mathcal{R}\), and complex conjugation \(\mathcal{C}\), and will henceforth be referred to as \(\mathcal{RC}\)-symmetry.

Any operator (such as \(\widehat K^2\) mapping between the spectra of real fields thus obeys the symmetry
\begin{equation}
    \mathcal{RC}\widehat K^2\mathcal{RC}=\widehat K^2.
    \label{eq:RCOperator}
\end{equation}
By applying this relation to the eigenvalue problem \(k^2\left|k^2\right\rangle=\widehat K^2\left|k^2\right\rangle\), eigenvectors and eigenvalues are found to fall into at least one of the classes~\cite{Hooper2025, Bender1998, Bender2007}:

\begin{enumerate}
    \label{list:RCSymmetryBreakingClassification}
    \item Symmetry-Broken Pairs: \({\rm{Im}}\left\{k^2\right\}\neq0\). There exists another eigenvalue \(\kappa^2=\left(k^2\right)^*\).
    \label{item:RCBrokenModes}
    \item Symmetry-Unbroken Modes:
    \({\rm{Im}}\left\{k^2\right\}=0\).
    \(\left|k^2\right\rangle\) can be chosen such that \(\mathcal{RC}\left|k^2\right\rangle=\left|k^2\right\rangle\): the eigenvector corresponds to a real-valued time-domain field. Such ``\(\mathcal{RC}\)-symmetric'' waves are unique to time-varying media~\cite{Hooper2025}.
    \label{item:RCUnbrokenMode}
\end{enumerate}

As the only continuous path from one class to another involves the collision of a pair of eigenvalues, the space of \(\mathcal{RC}\)-symmetric operators is thus divided into topologically distinct regions.
Thus, whilst this letter focuses on spatial adiabatic propagation, our results apply to any adiabatic evolution determined by an \(\mathcal{RC}\)-symmetric operator.

Having identified topological behaviour, we now focus on its observation, via the existence of a ``partially quantised'' complex Berry phase, as an \(\mathcal{RC}\)-symmetric wave adiabatically evolves under Eq. \ref{eq:OperatorWaveEquation} for \(\widehat K^2\left(z/L\right)\) following a closed loop in parameter space.

Although explicit calculation of the Berry phase may involve inconvenient integrals, its intuitive meaning is straightforward: it is simply the geometric phase (independent of \(L\)) evolved by a wave propagating through a system following a closed loop in parameter space. As such, in all to follow, we take \(\widehat K^2\left(0\right)=\widehat K^2\left(1\right)\), and follow the adiabatic propagation of an \(\mathcal{RC}\)-symmetric (right) eigenvector \(\Ket{0}_{\rm RC}\) of \(\widehat K^2\left(0\right)\). This is given by (see Appendix \ref{app:AdiabaticityDerivation})

\begin{equation}
    \widetilde E_{\rm ad}\left(z\right) = \sqrt{\frac{k\left(0\right)}{k\left(z\right)}}{\rm e}^{{\rm i}\Delta\phi_{\rm Berry,RC}\left(z\right)}
    \cos{\left[\phi_{\rm dyn}\left(z\right)\right]}\Ket{z,{\rm RC}}.
    \label{eq:AdiabaticDisplacementEvolution}
\end{equation}

Here, \(\Ket{z,{\rm RC}}\) is taken as the (continuously-varying) instantaneous \(\mathcal{RC}\)-symmetric eigenvector of the operator \(\widehat K^2\left(z/L\right)\), with a corresponding positive eigenvalue \(k^2\left(z\right)\). Under the adiabatic approximation, the effects of propagation simply serve to multiply this eigenvector by a scalar, which separates into \(3\) terms: \(\sqrt{\frac{k\left(0\right)}{k\left(z\right)}}\) is the geometric term associated with the WKB approximation; \(\exp\left[{{\rm i}\Delta\phi_{\rm Berry,RC}\left(z\right)}\right]\) contains the Berry phase associated with the changing eigenvector \(\Ket{z,{\rm RC}}\); and \(\cos{\left[\phi_{\rm dyn}\left(z\right)\right]}\) corresponds to the standing wave in space typically of \(\mathcal{RC}\)-symmetric waves (as noted in~\cite{Hooper2025}).

\begin{figure}
    \centering
    \includegraphics[width=0.8\linewidth]{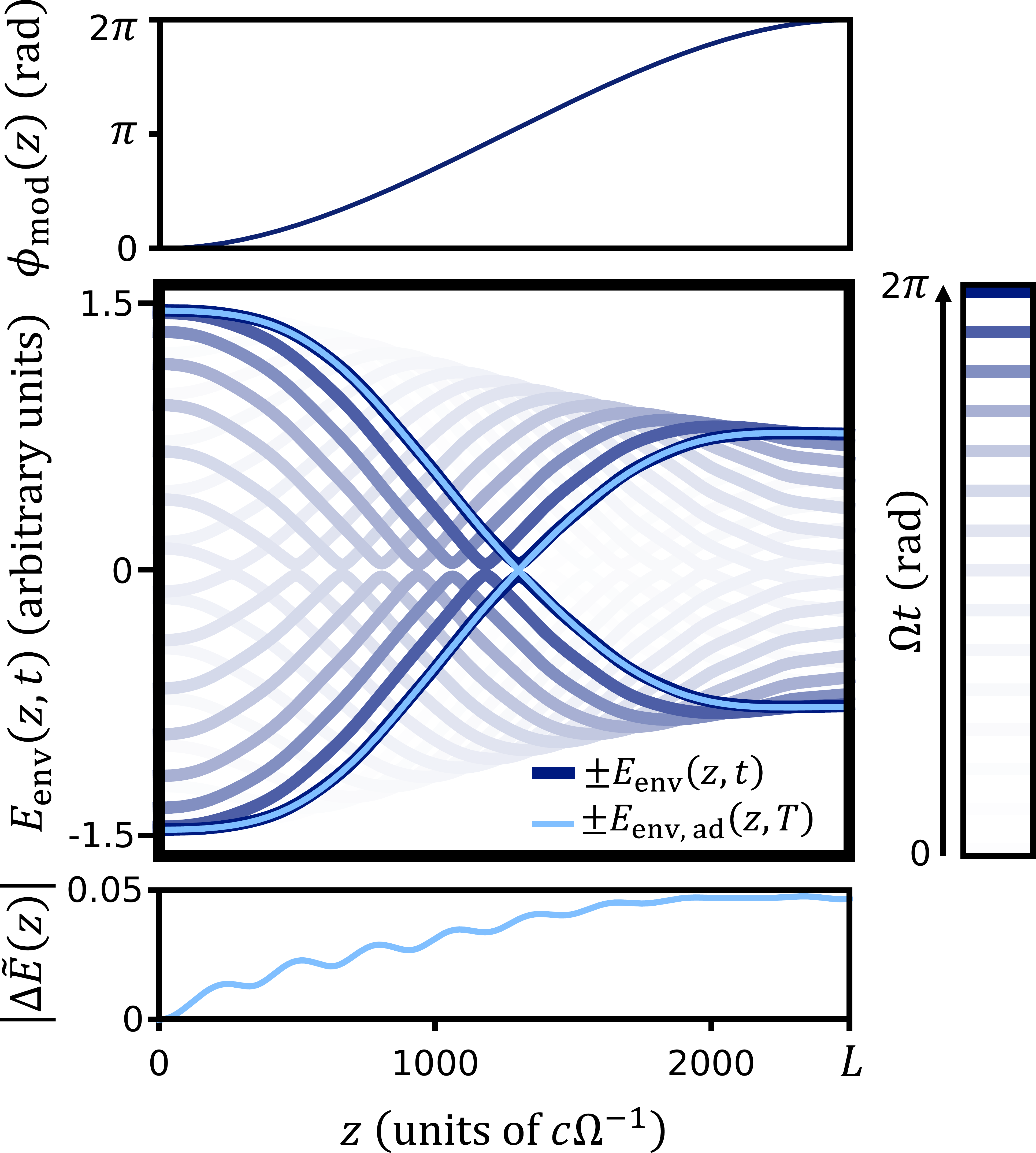}
    \caption{
        \textbf{The envelope enclosing a topological wave and its adiabatic approximation:}
        This example considers waves evolving from the left-hand side of a time-crystal with modulation frequency \(\Omega\), and a modulation phase \(\phi_{\rm mod}\left(z\right)\) varying slowly in space (top panel). The precise parameters are the same as ~\cite{Hooper2025, Hooper2026}.
        Middle panel: The numerically calculated envelope evolving over time (thick lines), and the corresponding adiabatic approximation at the final time (thinner line, overlaid) of an \(\mathcal{RC}\)-symmetric wave in a large cavity, with time-evolution pictured as a fading after-image.
        Geometric loss is demonstrated by the reduction in the height of the wave across the cavity, whilst a \(\pi\) geometric phase is revealed by the node in the adiabatic approximation. This does not form a true node in the numerical solution, due to scattering to frequencies outside of the adiabatic approximation, decreasing to \(0\) as \(L\to\infty\).
        Bottom panel: The total error in the spectra of the adiabatic approximation as a function of position, in the same units as the middle panel, demonstrating the accuracy of the adiabatic approximation.
    }
    \label{fig:ConvergenceDemo}
\end{figure}

This approximation is plotted, demonstrating its accuracy for large \(L\), in Fig. \ref{fig:ConvergenceDemo}, with the remaining phases given explicitly by

\begin{align}
    \phi_{\rm dyn}\left(z\right) &= \int_0^z k\left(z\right) {\rm d}z, \label{eq:DynamicalPhase} \\
    \Delta\phi_{\rm Berry,RC}\left(z\right) &= \int_0^z \Bra{z,{\rm RC}} {\rm i}\frac{\rm d}{{\rm d}z}\Ket{z,{\rm RC}} {\rm d}z, \label{eq:RCBerryPhase}
\end{align}
where the left eigenvector \(\Bra{z,{\rm RC}}\) is normalised such that \(\Braket{z,{\rm RC}|z,{\rm RC}}=1\). Although we present these results for \(\mathcal{RC}\)-symmetric eigenvectors (as a convenient and natural generalisation of the parallel transport gauge), the Berry phase maintains its gauge-invariance from quantum mechanics, with the derivation of the geometric phase in Appendix \ref{app:AdiabaticityDerivation} holding in any differentiable gauge.

Since preservation of \(\mathcal{RC}\)-symmetry is an incredibly important feature of Eq. \ref{eq:OperatorWaveEquation}, and with the adiabatic approximation becoming increasingly accurate for large \(L\), it is necessary that \(\widetilde\Psi_{\rm ad}\left(\xi;L\right)\) must come arbitrarily close to realising this symmetry for large \(L\). Thus inspired, we explicitly find the \(\mathcal{RC}\)-conjugate of (\ref{eq:AdiabaticDisplacementEvolution}):

\begin{align}
    \mathcal{RC}\widetilde E_{\rm ad}\left(z\right) = & \sqrt{\frac{k\left(0\right)}{k\left(z\right)}}
    \left({\rm e}^{{\rm i}\Delta\phi_{\rm Berry,RC}\left(\xi\right)}\right)^*\notag\\&\cos{\left[\phi_{\rm dyn}\left(z\right)\right]}\Ket{z,{\rm RC}}.
    \label{eq:ConjugatedAdiabaticEvolution}
\end{align}

Since \(\Delta\phi_{\rm Berry,RC}\left(z\right)\) is entirely independent of \(L\) (by Eq. \ref{eq:RCBerryPhase}), this symmetry can only hold as \(L\to\infty\) if \({\rm Im}\left\{\exp\left[{{\rm i}\Delta\phi_{\rm Berry,RC}\left(z\right)}\right]\right\}=0\). Furthermore, since \(\Delta\phi_{\rm Berry,RC}\left(0\right)=0\), before varying continuously with \(z\), this may be strengthened to \({\rm Re}\left\{\Delta\phi_{\rm Berry,RC}\left(z\right)\right\}=0\).

However, this term does not encode the full phase difference between \(\widetilde E_{\rm ad}\left(0\right)\) and \(\widetilde E_{\rm ad}\left(L\right)\). Despite the fact that \(\Ket{0,{\rm RC}}\) and \(\Ket{L,{\rm RC}}\) are both eigenvectors of \(\widehat K^2\left(0\right) = \widehat K^2\left(L\right)\) with the same eigenvalue, the requirement that \(\Ket{z,{\rm RC}}\) be differentiable for all \(z\) may constrain them such that \(\Braket{0,{\rm RC}|L,{\rm RC}}\neq 1\) (an example is later given by Eq. \ref{eq:ShiftedOddEigenvector}). However, even if this is the case, the fact that both correspond to real time-domain fields requires that \(\rm{Im}\left\{\Braket{0,{\rm RC}|L,{\rm RC}}\right\}=0\).

Thus, the total geometric phase difference is given by
\begin{equation}
    \Phi_{\rm Berry} = \Delta\phi_{\rm Berry,RC}\left(L\right)-{\rm i} \ln\left(\Braket{0,{\rm RC}|L,{\rm RC}}\right),
    \label{eq:TotalRCBerryPhase}
\end{equation}
resulting in the partial quantisation condition
\begin{equation}
    {\rm Re}\left\{\Phi_{\rm Berry}\right\} = 0\pmod{\pi}.
    \label{eq:PartialQuantisationCondition}
\end{equation}

In contrast to the familiar Berry phase from quantum mechanics, \(\Phi_{\rm Berry}\) is no longer a real number. As a result, Eq. \ref{eq:PartialQuantisationCondition} no longer represents a complete quantisation of the Berry phase onto a discrete set of points, instead presenting an analogous non-Hermitian generalisation: its partial quantisation onto lines of constant real parts.

As the simplest non-trivial example, consider a photonic time crystal with modulation frequency \(\Omega>0\), where coupling only occurs between the two frequencies \(\pm\frac{\Omega}{2}\). Within this space, \(\widehat K^2\left(z\right)\) takes the form~\cite{Hooper2025}

\begin{equation}
    \widehat K^2\left(z\right)=\left(\frac{\Omega}{2c}\right)^2
    \begin{pmatrix}
        1+\chi_0^*\left(z\right) & \chi_1\left(z\right) \\
        \chi_1^*\left(z\right) & 1+\chi_0\left(z\right)
    \end{pmatrix},
    \label{eq:2By2K2Example}
\end{equation}
where \(\chi_0\) encodes the average susceptibility at a frequency \(\Omega/2\), whilst \(\chi_1\) encodes the scale of the temporal modulation. A key feature of note in this system is that shifting the modulation by an arbitrarily phase results in a shift in the phase of to \(\chi_1\). This is in stark contrast to the phase of \(\chi_0\), which has far more direct physical implications, encoding the intrinsic static gain or loss in the system.

\begin{figure}
    \centering
    \includegraphics[width=0.5\linewidth]{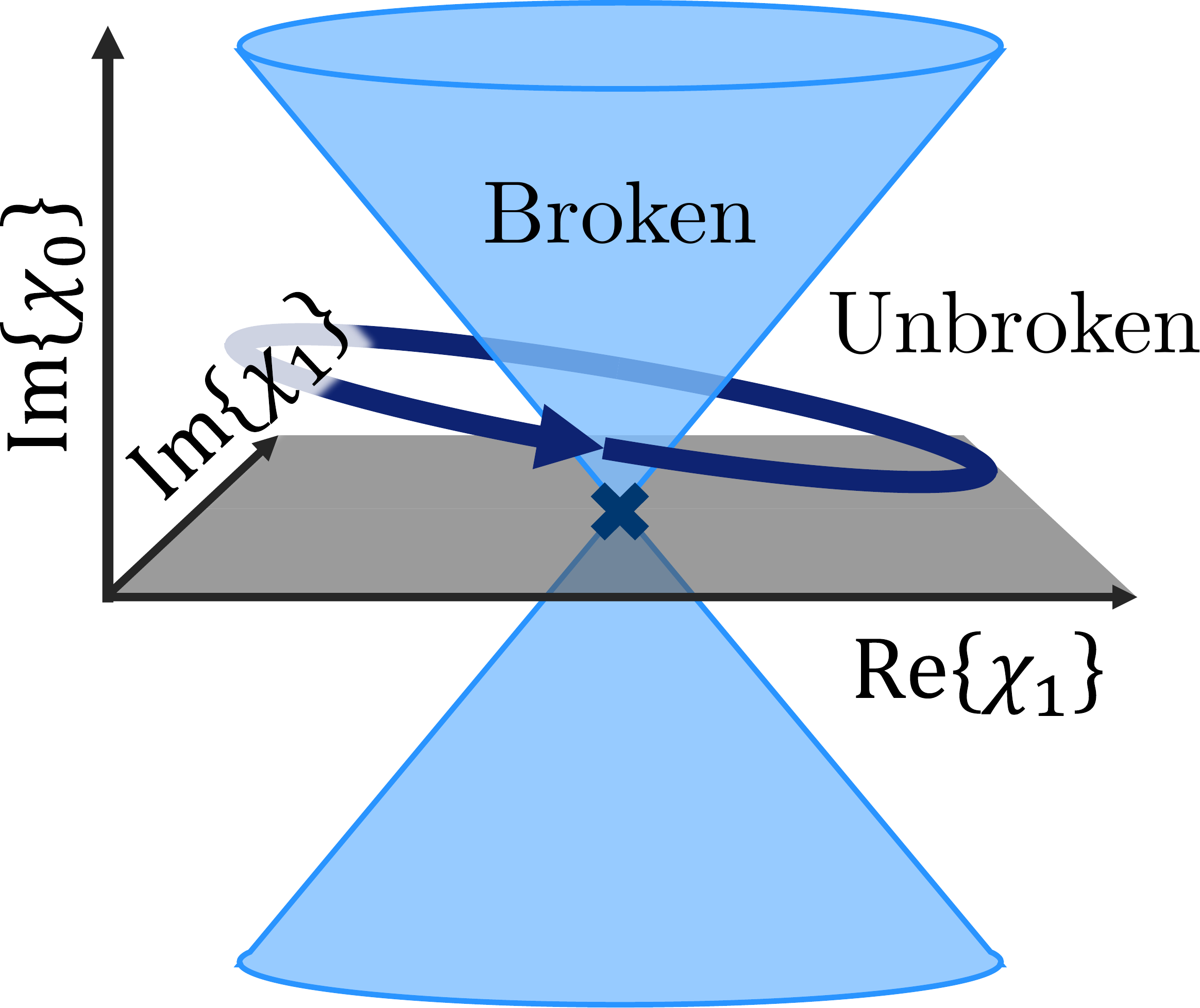}
    \caption{
        \textbf{Adiabatically distinct regions of broken and unbroken symmetry in the parameter space of a \(2\times2\) \(\mathcal{RC}\)-symmetric operator:}
        Explicit analysis of Eq. \ref{eq:2By2K2Example} reveals that eigenvalue collisions in the \(2\times2\) square wavenumber operator are determined entirely by Fourier components of the susceptibility. In particular, symmetry-breaking occurs when the average loss or gain \({\rm Im}\left\{\chi_0\right\}\) exceeds its time-variation \(\left|\chi_1\right|\), as is the case in all static media.
        Plotted, these regions form the interior of a double-cone, in which all loops in parameter space are contractible. By contrast, there exist non-contractible loops in regions of unbroken symmetry. Quantisation of \({\rm Re}\left\{\Phi_{\rm Berry}\right\}\) is thus determined solely by a winding number around the double cone.
        }
    \label{fig:DoubleCone}
\end{figure}

A straightforward calculation predicts that collisions between eigenvalues occur when
\begin{equation}
    {\rm Im}\left\{\chi_0\right\}=\pm \left|\chi_1\right|.
    \label{eq:2By2SymmetryBreakingCondition}
\end{equation}
Thus, the adiabatically distinct regions in the parameter space of \(\widehat{K}^2\) may be understood simply in terms of the variables
\(\begin{pmatrix}{\rm Re}\left\{\chi_1\right\} & {\rm Im}\left\{\chi_1\right\} & {\rm Im}\left\{\chi_0\right\}\end{pmatrix}\in\mathbb R^3\), forming the interior(exterior) of a double cone, corresponding to waves with spontaneously broken(unbroken) \(\mathcal{RC}\)-symmetry (see Fig. \ref{fig:DoubleCone}).

Since a continuous deformation of an adiabatic loop in parameter space must result in a continuous change to the Berry phase, any contractible loop must obey \({\rm Re}\left\{\Phi_{\rm Berry}\right\}=0\).

Following this continuity argument, determining the relationship between winding number and Berry phase requires only a single example for each winding number \(n\). To this end, we consider the example of \(\chi_0=0\), and \(\chi_1=\exp\left(2n\pi{\rm i} z/L\right)\). \(\widehat K^2\left(z\right)\) is then Hermitian, with eigenvectors given by
\begin{equation}
    \Ket{z}_{\rm RC}=
    \begin{pmatrix}
        \sqrt{\pm1}\exp\left({n\pi {\rm i} z/L}\right)
        \\
        \pm\sqrt{\pm1}\exp\left(-{n\pi {\rm i} z/L}\right)
    \end{pmatrix}.
    \label{eq:2By2EigenvectorExample}
\end{equation}

Substituting these eigenvectors into to Eq. \ref{eq:TotalRCBerryPhase}, the direct relation between winding number and Berry phase becomes
\begin{equation}
    {\rm Re}\left\{\Phi_{{\rm Berry},n}\right\} = n\pi\pmod{2\pi}.
\end{equation}

However, explicit calculations are not limited to finite dimensional examples. Instead, consider a time-varying system where the phase of some general periodic modulation with frequency \(\Omega\), slowly shifts in space by a time \(\Delta t=\frac{2n\pi}{\Omega}\frac{z}{L}\). Evaluating the Berry phase in this system is made tractable by noting that the result of shifting the modulation in time must result in equivalent shifts in eigenvectors.

To make this shift explicit, consider the \(\mathcal{RC}\)-symmetric eigenvectors of \(\widehat K^2\left(z=0\right)\) for a particular eigenvalue \(k^2\), where \(\Ket\omega_\mathcal{F}\) denotes a Fourier component \({\rm e}^{-{\rm i}\omega t}\) in the time-domain. These are naturally split into two categories: those containing frequencies at integer multiples of \(\Omega\) (henceforth referred to as even eigenvectors)
\begin{align}
    \Ket{k^2;z=0}_{\rm Even}=&A_0\Ket{0}_\mathcal{F}\notag\\+&\sum_{m=1}^\infty\left(A_m\Ket{m\Omega}_\mathcal{F}+A_m^*\Ket{-m\Omega}_\mathcal{F}\right),
    \label{eq:EvenEigenvectorAnsatz}
\end{align}
with \(A_0\in\mathbb R\); and those containing frequencies at odd-half integer multiples of \(\Omega\) (henceforth referred to as odd eigenvectors)
\begin{align}
    \Ket{k^2;z=0}_{\rm Odd}=&
    \sum_{m=1}^\infty B_m\Ket{\left(m-1/2\right)\Omega}_\mathcal{F}\notag\\
    +&\sum_{m=1}^\infty B_m^*\Ket{-\left(m-1/2\right)\Omega}_\mathcal{F}.
    \label{eq:OddEigenvectorAnsatz}
\end{align}

With these ansätze, the result of shifting the modulation by a time \(\Delta t=\frac{2\pi}{\Omega}\frac{z}{L}\) becomes explicit. In the even case,
\begin{align}
    \Ket{k^2;z}_{\rm Even}&=A_0\Ket{0}_\mathcal{F}\notag\\
    &+\sum_{m=1}^\infty A_m{\rm e}^{2mn{\rm i}\pi z/L}\Ket{m\Omega}_\mathcal{F}\notag\\&+\sum_{m=1}^\infty A_m^*{\rm e}^{-2mn{\rm i}\pi z/L}\Ket{-m\Omega}_\mathcal{F},
    \label{eq:ShiftedEvenEigenvector}
\end{align}
and in the odd case,

\begin{align}
    \Ket{k^2;z}_{\rm Odd}=&
    \sum_{m=1}^\infty B_m{\rm e}^{\left(2m-1\right)n{\rm i}\pi z/L}\Ket{\left(m-1/2\right)\Omega}_\mathcal{F}\notag\\
    +&\sum_{m=1}^\infty B_m^*{\rm e}^{-\left(2m-1\right)n{\rm i}\pi z/L}\Ket{-\left(m-1/2\right)\Omega}_\mathcal{F}.
    \label{eq:ShiftedOddEigenvector}
\end{align}

One may then immediately verify that \(\Ket{k^2;0}_{\rm Even}=\Ket{k^2;L}_{\rm Even}\), and \(\Ket{k^2;0}_{\rm Odd}=\left(-1\right)^n\Ket{k^2;L}_{\rm Odd}\), which may then be substituted into Eq. \ref{eq:TotalRCBerryPhase}, to find the general partial quantisations
\begin{subequations}
    \label{eq:PartialQuantisationWithShifts}
    \begin{align}
        {\rm Re}\left\{\Phi_{\rm Berry,Even}\right\}&=0 \pmod{2\pi},
        \label{subeq:EvenPartialQuantisation}\\
        {\rm Re}\left\{\Phi_{\rm Berry,Odd}\right\}&=n\pi \pmod{2\pi},
        \label{subeq:OddPartialQuantisation}
    \end{align}
\end{subequations}
which, as previously, characterise not only these particular trajectories, but any adiabatically connected to them.

The loss in intuition as familiar quantities fail to commute is not a difficulty unique to time-varying media, and nor are its particular complexities disjoint from all other fields: tools from general relativity~\cite{Horsley2023BlackHole} and quantum mechanics~\cite{Horsley2023Operators} have already been applied to solve time-varying problems. Thus, this letter considered how ideas of topology originating from quantum mechanics translate to the non-Hermitian setting of time-varying media. We found that the real-valued nature of classical waves guarantees that the operators describing their propagation possess sufficient symmetry to realise non-trivial topology, and we described how this topology may be characterised by a simple Berry phase measurement.

Nor are these results limited to the context of spatial propagation in which they were presented. Time-domain analyses~\cite{Hooper2026, Valero2025, Vial2025, MartiSabate2025} obey the same \(\mathcal{RC}\)-symmetry~\cite{Hooper2026}, raising the possibility of observing topological effects in even a single time-varying resonator~\cite{IHooper2025}, with a nearly periodic, but slowly drifting, modulation (see Appendix \ref{app:FQNMAdiabaticity}). This generality should make these results far easier to apply in experimental settings.

Finally, we note that, whilst we have identified an important symmetry and a number of experimentally accessible topological consequences, we have not attempted to present an exhaustive characterisation of this topology. Thus, a promising avenue for future studies is to consider the range of existing work on non-Hermitian topology~\cite{Okuma2023, Kawabata2026, Gong2018, Borgnia2020}, and to see which of those phenomena have analogues protected by \(\mathcal{RC}\)-symmetry, and may thus be realised in time-varying media.

\begin{acknowledgments}
CMH thanks Simon Horsley, Ian Hooper, Oliver Breach, and James Walkling for useful discussions, and acknowledges financial support from the Engineering and Physical Sciences Research Council (EPSRC) of the UK via the Exeter University Physics DTP, and via the META4D Programme Grant
(EP/Y015673/1).
\end{acknowledgments}

\bibliography{apssamp}

\appendix

\section{\label{app:AdiabaticityDerivation}The Adiabatic Approximation for Lossy Wave Propagation}
This section contains a quick derivation of Eq. \ref{eq:AdiabaticDisplacementEvolution}. To begin, note that Eq. \ref{eq:OperatorWaveEquation} may be written, with \(\xi=z/L\) in matrix form as
\begin{equation}
    \frac{{\rm d}}{{\rm d}\xi}
    \begin{pmatrix}
        \widetilde\Psi\left(\xi\right) \\ \partial_\xi\widetilde\Psi\left(\xi\right)
    \end{pmatrix}
    =
    \begin{pmatrix}
        0 & 1 \\
        -\widehat K^2\left(\xi\right)L^2 & 0
    \end{pmatrix}
    \begin{pmatrix}
        \widetilde\Psi\left(\xi\right) \\ \partial_\xi\widetilde\Psi\left(\xi\right)
    \end{pmatrix}.
    \label{eq:FirstOrderWaveEquation}
\end{equation}

Our focus in this section will then be directed towards approximating an evolution operator for the above equation, given by

\begin{align}
    \frac{{\rm d} \widehat U\left(\xi;0\right)}{{\rm d}\xi}
    =&
    \begin{pmatrix}
        0 & 1 \\
        -\widehat K^2\left(\xi\right)L^2 & 0
    \end{pmatrix}
    \widehat U\left(\xi;0\right), \\
    \widehat U\left(0;0\right) = 1.
    \label{eq:FirstOrderPropagatorEquation}
\end{align}

The adiabatic approximation will proceed by considering the instantaneous diagonalisation
\begin{align}
    \begin{pmatrix}
        0 & 1 \\
        -\widehat K^2\left(\xi\right)L^2 & 0
    \end{pmatrix}=&
    \left[\widehat X_{\rm W}\left(\xi\right)\widehat X_K\left(\xi\right)\right]\notag \\
    &\cdot
    \begin{pmatrix}
        {\rm i}\widehat D_KL\left(\xi\right) & 0 \\
        0 & -{\rm i}\widehat D_KL\left(\xi\right)
    \end{pmatrix}\notag \\
    &\cdot
    \left[\widehat X_K^{-1}\left(\xi\right)\widehat X_{\rm W}^{-1}\left(\xi\right)\right],
    \label{eq:CompleteWaveDiagonalisation}
\end{align}
where \(\widehat X_K\left(\xi\right) \widehat D_K(\xi)\widehat X_K^{-1}\left(\xi\right)=\widehat K\left(\xi\right)\), and
\begin{equation}
    \widehat X_{\rm W}\left(\xi\right)=
    \begin{pmatrix}
        1 & 0 \\
        0 & {\rm i} \widehat K\left(\xi\right)L
    \end{pmatrix}
    \begin{pmatrix}
    1 & 1 \\ 1 & -1
    \end{pmatrix}.
    \label{eq:WaveDiagonalisation}
\end{equation}

It is then straightforward to find that
\begin{align}
    \frac{{\rm d}\widetilde U\left(\xi;0\right)}{{\rm d}\xi}\notag = \\ 
    & \begin{pmatrix}
        {\rm i}\widehat D_KL\left(\xi\right) & 0 \\
        0 & -{\rm i}\widehat D_KL\left(\xi\right)
    \end{pmatrix}\widetilde U\left(\xi;0\right)\notag\\
    & -\widehat X_K^{-1}\left(\xi\right)\widehat X_{\rm W}^{-1}\left(\xi\right) \frac{{\rm d}\widehat X_{\rm W}\left(\xi\right)}{{\rm d}\xi}\widehat X_K\left(\xi\right)\widetilde U\left(\xi;0\right)\notag\\
    & -\widehat X_K^{-1}\left(\xi\right) \frac{{\rm d}\widehat X_K\left(\xi\right)}{{\rm d}\xi}\widetilde U\left(\xi;0\right),
    \label{eq:LocallyDiagonalisedPropagatorEquation}
\end{align}
where
\begin{equation}
    \widetilde U\left(\xi;0\right)=\widehat X_K^{-1}\left(\xi\right)\widehat X_{\rm W}^{-1}\left(\xi\right)\widehat U\left(\xi;0\right)\widehat X_K\left(0\right)\widehat X_{\rm W}\left(0\right).
    \label{eq:LocallyDiagonalisedPropagator}
\end{equation}

The adiabatic approximation then arises when one neglects the off-diagonal components on the coefficients of \(\widetilde U\left(\xi;0\right)\). To do so gives rise the following equation:

\begin{align}
    \frac{{\rm d}\widetilde U_{\rm ad}\left(\xi;0\right)}{{\rm d}\xi}\notag = \\ 
    & \begin{pmatrix}
        {\rm i}\widehat D_KL\left(\xi\right) & 0 \\
        0 & -{\rm i}\widehat D_KL\left(\xi\right)
    \end{pmatrix}\widetilde U_{\rm ad}\left(\xi;0\right)\notag\\
    & +\widehat D_{\rm WKB}\left(\xi\right)\widetilde U_{\rm ad}\left(\xi;0\right)\notag\\
    & +{\rm i}\widehat D_{\rm Berry}\left(\xi\right) U_{\rm ad}\left(\xi;0\right),\notag\\
    \widetilde U_{\rm ad}\left(0;0\right) =& 1,
    \label{eq:LocallyDiagonalisedAdiabaticPropagatorEquation}
\end{align}
where
\begin{align}
    \widehat D_{\rm WKB}\left(\xi\right)=&-\frac{1}{2}\frac{{\rm d} \ln\left(\widehat D_K(\xi)\right)}{{\rm d}\xi},
    \label{eq:WKBOperator}
    \\
    \widehat D_{\rm Berry}\left(\xi\right)=&{\rm ondiag}\left(\widehat X_K^{-1}\left(\xi\right) {\rm i}\frac{{\rm d}}{{\rm d}\xi}\widehat X_K\left(\xi\right)\right).
    \label{eq:BerryPhaseOperator}
\end{align}

This equation may be simply solved for \(\widetilde U_{\rm ad}\left(\xi;0\right)\) to give

\begin{align}
    \widetilde U_{\rm ad}\left(\xi;0\right) =&
    \widetilde U_{\rm WKB}\left(\xi;0\right)
    \widetilde U_{\rm Berry}\left(\xi;0\right)\\&\cdot
    \begin{pmatrix}
        \widetilde U_{\rm dyn,+}\left(\xi;0\right) & 0 \\
        0 & \widetilde U_{\rm dyn,-}\left(\xi;0\right)
    \end{pmatrix},
    \label{eq:LocallyDiagonalisedAdiabaticPropagatorSolution}
\end{align}
where
\begin{align}
    \widetilde U_{\rm dyn,\pm}\left(\xi;0\right) =& \exp\left[\pm{\rm i}L\int_0^\xi \widehat D_K\left(\xi'\right){\rm d}\xi'\right], \\
    \widetilde U_{\rm WKB}\left(\xi;0\right) =& \left(\widehat D_K\left(0\right)\right)^{1/2}\left(\widehat D_K\left(\xi\right)\right)^{-1/2}, \\
    \widetilde U_{\rm Berry}\left(\xi;0\right) =& \exp\left[{\rm i}\int_0^\xi \widehat D_{\rm Berry}\left(\xi'\right){\rm d}\xi'\right].
    \label{eq:LocallyDiagonalisedAdiabaticPropagatorComponents}
\end{align}

Since \(\widetilde U_{\rm ad}\left(\xi;0\right)\) approximates \(\widetilde U\left(\xi;0\right)\), by inverting Eq. \ref{eq:LocallyDiagonalisedPropagator} one immediately finds the approximation \(\widehat U_{\rm ad}\left(\xi;0\right)\) to \(\widehat U\left(\xi;0\right)\). The result presented in Eq. \ref{eq:AdiabaticDisplacementEvolution} is found by evaluating

\begin{equation}
    \widehat \Psi_{\rm ad}\left(\xi;L\right)=
    \begin{pmatrix}
        1 \\ 0
    \end{pmatrix}\cdot \widehat U_{\rm ad}\left(\xi;0\right)
    \begin{pmatrix}
        1 \\ 0
    \end{pmatrix}
    \frac{1}{\sqrt{k\left(0\right)}}\Ket{0,{\rm RC}}.
    \label{eq:InitiallyFlatAdiabaticExample}
\end{equation}

\section{\label{app:FQNMAdiabaticity}Adiabaticity in Floquet Quasinormal Mode Problems}

Consider the evolution of a wave within a cavity. In general, this may be phrased as
\begin{equation}
    \left(\frac{\rm d}{{\rm d}t}+M\left(t,\tau\right)\right)\Psi\left(t\right)=0,
    \label{eq:CavityEvolutionProblem}
\end{equation}
where \(\Psi\left(t\right)\) is a Hilbert space vector describing the mode of a cavity as a function of a time \(t\), while \(M\left(t,\tau\right)\) is an operator acting on this space to determine its time-evolution, written explicitly in terms of a time-variable \(t\), and a slowly-varying parameter \(\tau=\epsilon t\), for \(\epsilon \ll 1\).

Into this, one may substitute the ansatz
\begin{equation}
    \Psi\left(t\right)=\int_{-\infty}^\infty \widetilde\Psi\left(\omega,\tau\right){\rm e}^{-{\rm i}\omega t} \frac{{\rm d}\omega}{2\pi}.
    \label{eq:FourierAnsatz}
\end{equation}

Noting that \(\frac{\rm d}{{\rm d}t}\) may be written formally as \(\frac{\partial}{\partial t}+\epsilon \frac{\partial}{\partial t}\), and following the techniques of~\cite{Horsley2023Operators}, Eq. \ref{eq:CavityEvolutionProblem} may be rewritten in the form
\begin{equation}
    \left(\frac{\partial}{\partial \tau}+\epsilon^{-1}\left(-{\rm i}\widehat\omega+\widehat M\left(\tau\right)\right)\right)\Ket{\widetilde\Psi\left(\tau\right)}=0,
    \label{eq:MultipleScalesFourierEvolution}
\end{equation}
where \(\widehat M\) and \(\widehat \omega\) both act upon the spectrum vector \(\Ket{\widetilde \Psi\left(\tau\right)}\). Note that, despite following the notation of multiple scales analysis, this rearrangement is exact.

Recalling the condition that \(\epsilon\ll1\), the adiabatic approximation then provides a good description of this evolution in terms of the instantaneous eigenmodes
\begin{equation}
    \left(-{\rm i}\widehat\omega+\widehat M\left(\tau\right)\right)\Ket{\widetilde\Psi\left(\tau\right)}_{\rm ins}={\rm i}\omega_0\Ket{\widetilde\Psi\left(\tau\right)}_{\rm ins},
    \label{eq:InstantaneousEigenvectors}
\end{equation}
which are precisely the modes of the system when \(\tau\) is taken as a constant. Since \(M\left(t,\tau\right)\) and \(\frac{\partial}{\partial t}\) map between real-valued fields in the time-domain, it follows that \(\widehat M\) and \({\rm i}\widehat\omega\) obey \(\mathcal{RC}\)-symmetry, and thus that the partial quantisation derived in this letter applies whenever an \(\mathcal{RC}\)-unbroken \(\Ket{\widetilde\Psi\left(\tau\right)}_{\rm ins}\) is inserted into the system. In the case of nearly-periodic media, \(M\left(t,\tau\right)\)=\(M\left(t+\frac{2\pi}{\Omega},\tau\right)\), and the instantaneous eigenmodes become precisely the Floquet quasinormal modes studied in~\cite{Hooper2026, Valero2025, Vial2025, MartiSabate2025}. Furthermore, it was demonstrated in~\cite{Hooper2026} that \(\mathcal{RC}\)-symmetric modes are ubiquitous in such systems, providing a practical alternative to achieving \(\mathcal{RC}\)-unbroken modes in \(\widehat K^2\).

\end{document}